\documentclass{iau} 
\usepackage{graphicx}
\usepackage{natbib}
\usepackage{url}

\def\vector#1{\mbox{\boldmath $#1$}}

\newcommand{\kpc}{\ensuremath{\,\mathrm{kpc}}}
\newcommand{\Myr}{\ensuremath{\,\mathrm{Myr}}}

\newcommand{\mas}{\ensuremath{\,\mathrm{mas}}}
\newcommand{\masyr}{\ensuremath{\,\mathrm{mas\ yr}^{-1}}}

\newcommand{\vlos}{v_{\ensuremath{\mathrm{los}}}}

\newcommand{\mualpha}{\mu_{\alpha*}}
\newcommand{\mudelta}{\mu_\delta}

\newcommand{\flag}[1]{\texttt{\lowercase{#1}}}
\newcommand{\agama}{\texttt{Agama}}
\newcommand{\emcee}{\flag{emcee}}
\newcommand{\Gaia}{\textit{Gaia}}

\title[Prolate Galactic dark matter halo] 
{The shape of the dark matter halo \\ revealed from a hypervelocity star }

\author[Hattori \& Valluri]   
{Kohei Hattori$^{1,2}$, Monica Valluri$^{1}$
}

\affiliation{$^1$Deptartment of Astronomy, University of Michigan, 
1085 S University Ave, Ann Arbor, MI 48109, USA.  email: {\tt khattori@umich.edu} \\[\affilskip]
$^2$Deptartment of Physics, Carnegie Mellon University, 
5000 Forbes Ave, Pittsburgh, PA 15213, USA}

\pubyear{2019}
\volume{353}  
\jname{Galactic Dynamics in the Era of Large Surveys}
\editors{M. Valluri \& J. A. Sellwood  }
\begin{document}

\maketitle

\begin{abstract}

A recently discovered young, high-velocity giant star J01020100-7122208 
is a good candidate of hypervelocity star ejected from the Galactic center, although it has a bound orbit.  
If we assume that this star was ejected from the Galactic center, 
it can be used to constrain the Galactic potential, 
because the deviation of its orbit from a purely radial orbit 
informs us of the torque 
that this star has received after its ejection. 
Based on this assumption, we estimate the flattening of the dark matter halo of the Milky Way 
by using the Gaia DR2 data and the circular velocity data from \citealt{Eilers2019}. 
Our Bayesian analysis shows that the orbit of J01020100-7122208 favors a prolate dark matter halo within $\sim$ 10 kpc from the Galactic center. 
The posterior distribution of the density flattening $q$ 
shows a broad distribution at $q \gtrsim1$ 
and peaks at  $q \simeq 1.5$.  
Also,  98.5\% of the posterior distribution is located at $q>1$, 
highly disfavoring an oblate halo.

\end{abstract}

\firstsection 

\section{Introduction}

Just after the discovery of the first hypervelocity star (HVS) candidate \citep{Brown2005}, 
\cite{Gnedin2005} proposed that the orbits of HVSs ejected from the Galactic center 
could be a used to constrain the Galactic potential (see also \citealt{Yu2007,Rossi2017}). 
They showed that the angle between the position and velocity vectors 
of a HVS in the Galactocentric frame 
gradually changes as it moves away from the Galactic center, 
due to the torque from the stellar disk and triaxial halo. 
This angle is typically $\sim 1^\circ$ or smaller, 
so a very accurate measurement of the position and velocity of a HVS is required to measure the shape of the dark matter distribution.

Now that we have more HVS candidates 
\citep{Brown2015, Bromley2018, Hattori2018a, Marchetti2018, Koposov2019}
and that reliable astrometric data are available from Gaia \citep{Gaia2018}, 
we can now apply their method to the data. 
Here, we use a recently discovered HVS candidate, J01020100-7122208 
(hereafter J0102), 
to estimate the dark halo's flattening $q$.  
We note that a HVS dubbed S5-HVS1 \citep{Koposov2019} moves too fast to constrain $q$, 
but it can constrain the Solar azimuthal velocity well with a method proposed by \cite{Hattori2018b}.

\section{Data}

\subsection{Data for the HVS candidate J0102 -- corrected for the systematic error}

J0102 was originally discovered by \cite{Neugent2018} as a high velocity star. 
Based on the Gaia DR2 data, the same group of authors \citep{Massey2018} confirmed that 
its orbit is consistent with a picture that this star was ejected from the Galactic center.

\cite{Massey2018} 
figured out that J0102 is a 3-4$M_\odot$, young G-type giant, whose age is estimated to be 180 Myr. 
The line-of-sight velocity of this star is $v_\mathrm{los} \pm \delta \vlos = 301 \pm 2.4 \;\mathrm{km\;s^{-1}}$. 
The metallicity is (roughly) estimated to be [Fe/H] $\simeq -0.5$.

The astrometric data for J0102 is available in Gaia DR2. 
(The source id for this star is Gaia DR2 4690790008835586304.) 
J0102 is a relatively bright star ($G=13.37$ mag) located at $(\alpha, \delta) = (15.5043, -71.3724)^\circ$, 
and its astrometric solution is well-behaved, judging from the modest value of RUWE$=0.993 (<1.4)$. 
The measured parallax is $\varpi_\mathrm{obs} \pm \delta\varpi_\mathrm{int}   =( 0.07276 \pm 0.01904) \mas$, 
and the measured proper motion is 
$(\mu_\mathrm{\alpha*, obs} \pm \delta \mu_\mathrm{\alpha*, int}, \mu_\mathrm{\delta, obs} \pm \delta \mu_\mathrm{\delta, int}) 
= (8.6465 \pm 0.03607,  -0.9062 \pm 0.02743) \masyr $. 
Here, we put the subscript `int' to stress that these uncertainties denote the Gaia's internal (formal) error reported in Gaia DR2. 
We neglect the small correlation coefficient ($=0.06993$) between the errors on proper motions.  
This does not affect our result much, because we will inflate the proper motion error to take into account the systematic error.

Following the presentation slides by L. Lindegren at IAU GA 30 (2018),\footnote{
\url{https://www.cosmos.esa.int/documents/29201/1770596/Lindegren_GaiaDR2_Astrometry_extended.pdf/}
}
we take into account the external (total) error including the systematic error.  
For the parallax, we adopt 
$\varpi  \pm \delta \varpi_\mathrm{ext} = (0.1017 \pm 0.04766)$ mas. 
Here, we add the zero-point offset of 0.029 mas to $\varpi_\mathrm{obs}$; 
and we use a formula to inflate the parallax error 
$\delta \varpi_\mathrm{ext} = [ (1.08 \times \delta\varpi_\mathrm{int})^2 + (0.043 \mas )^2 ]^{1/2}$. 
Similarly, we adopt 
$(\mu_\mathrm{\alpha*} \pm \delta \mu_\mathrm{\alpha *, ext}, \mu_\mathrm{\delta} \pm \delta \mu_\mathrm{\delta, ext})  = (8.6465 \pm 0.07663,  -0.9062 \pm 0.07234) \masyr $  
for the proper motion. 
Here, we use a formula to inflate the proper motion error 
$\delta \mu_\mathrm{ext} = [ (1.08 \times \delta\mu_\mathrm{int})^2 + (0.066 \masyr)^2 ]^{1/2}$.

\subsection{Data for the circular velocity curve}

It is hard to determine the global Galactic potential by using only a single HVS. 
Thus, we also use the circular velocity $v_c(R)$ from \cite{Eilers2019}. 
The random error on $v_c(R)$ is provided in their Table 1. 
We read off the approximate systematic error on $v_c$ from Figure 4  of \cite{Eilers2019}. 
We add the random and systematic errors on $v_c$ at each radius quadratically to estimate the total error $\delta v_c$, as in \cite{deSalas2019}.

\begin{figure}
\begin{center}
 \includegraphics[width=3.2in]{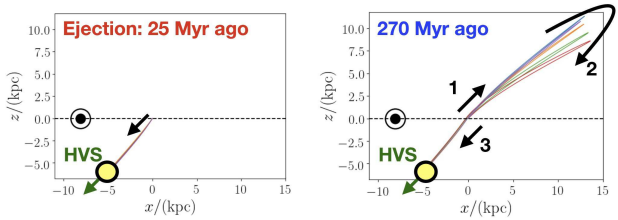} 
 \caption{Orbit of a bound HVS candidate J0102 viewed edge-on. 
 Randomly sampled orbits from our MCMC with different assumptions on the flight time are shown.}
   \label{fig1}
\end{center}
\end{figure}

\section{Formulation}

\subsection{Model potential of the Milky Way}

One of the best Galactic potential models is the one in \cite{McMillan2017}. 
In his model, the baryon potential $\Phi_\mathrm{baryon, M17}$ 
comprises of several components such as the bulge, thin/thick disks, and gas disk; 
and the dark matter halo is expressed  by a spherical NFW model. 
In our analysis, we adopt an axisymmetric potential model of the form 
$\Phi(R,z) = f_\mathrm{b} \Phi_\mathrm{baryon, M17} +  \Phi_\mathrm{DM}$. 
Here, $f_\mathrm{b} \sim 1$ is a  free parameter that controls the strength of the baryonic potential. 
(We note that it does not change the shape of the baryonic potential; it effectively changes the total baryonic mass.) 
The density profile of our dark matter halo model is given by a modified NFW profile of the form 
$\rho(R,z) = \rho_0 (m/a)^{-1} (1 + (m/a))^{-2}$, 
where $m^2 = R^2 + z^2/q^2$. 
With this parametrization, 
we have only a handful of parameters, $\vector{\Theta}_\Phi = (f_\mathrm{b}, a, q, \rho_0)$, for the potential 
while keeping the potential sufficiently realistic and flexible.

\subsection{Parametrization of the orbit}

We assume that J0102 was ejected from the Galactic center $t_\mathrm{flight}$ ago. 
Because the flight time $t_\mathrm{flight}$ cannot exceed the stellar age ($\sim 180 \Myr$ according to  \citealt{Massey2018}), 
we assume that $t_\mathrm{flight}<300\Myr$. 
Here, we make a conservative limit on $t_\mathrm{flight}$ 
so that any systematic error on the stellar age does not seriously affect our result (e.g., \citealt{Hattori2019}). 
Using the fiducial potential model and the point estimate of the 6D position of J0102, 
we check its orbit in the last 300 Myr. 
We find that there are two kinds of possible orbit for this star. 
The first is the `zero disc crossing' scenario, 
in which this star was ejected recently ($t_\mathrm{flight}<50 \Myr$)
and has never experienced disc crossing since then. 
The other is the `one disc crossing' scenario, 
in which the flight time has a moderate value of $50 \Myr < t_\mathrm{flight}< 300 \Myr$ 
and J0102 has experienced one disc crossing after the ejection. 
We do not know the number of disc crossing ($N_\mathrm{dc}$) 
that J0102 experienced after the ejection. 
Thus,  we introduce a random variable $\nu$, 
and probabilistically assign 
$N_\mathrm{dc}=0$ (`zero disc crossing') and $N_\mathrm{dc}=1$ (`one disc crossing') 
with equal weights. 
Simple prescriptions such as 
$N_\mathrm{dc} = \mathrm{int} (1+ \sin( 100 \pi \nu))$ work fine for this purpose.

We denote the current 6D position and velocity for J0102 as  
$\vector{u}^\mathrm{true} = (\varpi^\mathrm{true}, \alpha, \delta,  \vlos^\mathrm{true}, \mualpha^\mathrm{true}, \mudelta^\mathrm{true})$. 
Under the assumption that J0102 was ejected from the Galactic center, 
the quantities $(\vector{\Theta}_\Phi, \vector{u}^\mathrm{true}, \nu)$ 
must orchestrate such that the corresponding orbit goes through the Galactic center in the past. 
This means that these quantities cannot be varied freely, 
and we need to carefully design our Bayesian model. 
Hereafter, we select $(\vector{\Theta}_\Phi, \nu, \mualpha^\mathrm{true}, \mudelta^\mathrm{true})$
as the parameters in our Bayesian formulation. 
We treat 
$(\varpi^\mathrm{true},  \vlos^\mathrm{true})$ as intermediate variables that are  used only to compute the likelihood,  
and we treat $(\alpha, \delta)$ as constants.

\subsection{Bayesian formulation}

From Bayes' theorem, the probability distribution of the parameters 
$p =  \{ \vector{\Theta}_\Phi, \nu, \mualpha^\mathrm{true}, \mudelta^\mathrm{true}  \} $
given the data $D$ and the model $M$ 
(such as the functional form of the potential model and the error model) 
is expressed as  
$
\mathrm{Pr} (p | D, M) = 
{\mathrm{Pr} (D|p, M) \mathrm{Pr} (p)} / {\mathrm{Pr} (D| M)}. 
$

The prior $\mathrm{Pr} (p)$ for $(\vector{\Theta}_\Phi, \nu)$ is set as follows. 
We adopt flat priors for $-\infty < \nu <\infty$, $\log a$, and $\log \rho_0$. 
We introduce an auxiliary parameter $u=2/ \pi \arctan(q)$ \citep{Bowden2016,Posti2019}, 
and we adopt a flat prior of $0.1855471582 < u < 0.7951672353$. 
This range of $u$ corresponds to $0.3<q<3$. 
For $(f_\mathrm{b}, \mualpha^\mathrm{true}, \mudelta^\mathrm{true})$, 
we adopt Gaussian priors with mean $(1, \mualpha, \mudelta)$ and dispersion $(0.1, \delta \mu_\mathrm{\alpha *, ext}, \delta \mu_\mathrm{\delta, ext})$.

The likelihood $\mathrm{Pr} (D|p, M)$ is evaluated as follows. 
First, we transform $\nu$ to $N_\mathrm{dc}$. 
When $N_\mathrm{dc} = 0,1$, we set the allowed range of $t_\mathrm{flight}$ to be $0 < t_\mathrm{flight} < 50 \Myr$, 
$50 \Myr < t_\mathrm{flight} < 300 \Myr$, respectively. 
Secondly, under the gravitational potential given by $\vector{\Theta}_\Phi$, 
find a pair of $(\varpi^\mathrm{true},  \vlos^\mathrm{true})$ 
such that the orbit characterized by $\vector{u}^\mathrm{true}$ 
goes through the Galactic center and the flight time is within the allowed range. 
The likelihood is given by 
$\mathrm{Pr} (D|p, M) = L_\mathrm{HVS}  L_\mathrm{CV}$, 
where the contribution from the circular velocity is given by $L_\mathrm{CV}$ (e.g., \citealt{deSalas2019}). 
The contribution from J0102 is given by  
\begin{equation}
L_\mathrm{HVS} = 
\frac{1}{\sqrt{2 \pi \delta \varpi_\mathrm{ext}^2}} \exp \left[ - \frac{( \varpi - \varpi^\mathrm{true} )^2}{2 \delta \varpi_\mathrm{ext}^2 }\right] 
+ 
\frac{1}{\sqrt{2 \pi \delta \vlos^2}} \exp \left[ - \frac{( \vlos - \vlos^\mathrm{true} )^2}{2 \delta \vlos^2 }\right] . 
\end{equation}

The Bayesian evidence $\mathrm{Pr} (D| M)$ is treated as a constant.

\begin{figure}[h]
\begin{center}
 \includegraphics[width=3.2in]{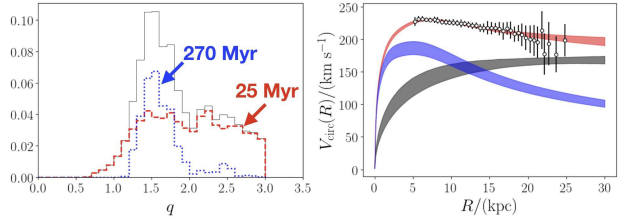} 
 \caption{Left: The posterior distribution of the flattening parameter $q$. 
 The contribution from orbits with different flight time are also shown. 
 Right: The circular velocity curve data from \cite{Eilers2019} along with our posterior distribution. 
 The shaded regions show the 68 percentile regions of the posterior distribution (red: total, blue: baryon, black: dark matter).} 
   \label{fig2}
\end{center}
\end{figure}

\section{Analysis and results}

We use a Markov Chain Monte Carlo package \emcee\  \citep{ForemanMackey2013} 
for our analysis. 
As in Fig. \ref{fig1}, J0102 can have two kinds of orbits. 
Left panel of Fig. \ref{fig2} shows the posterior distribution of $q$ 
and the contributions from the two kinds of orbits. 
We find that both orbits favor prolate halo (at $r \lesssim10 \kpc$). 
The total posterior peaks at $q\simeq 1.5$, 
and 98.5\% of the posterior  is located at $q>1$. 
At face value, our results agree with \cite{Posti2019}. 
However, we note that \cite{Posti2019} used \agama\ action finder \citep{Vasiliev2019AGAMA} for not only oblate system but also prolate system, 
which is mathematically invalid (see paragraph 4 of section 4.5 in \citealt{Vasiliev2019b}). 
Right panel of Fig. \ref{fig2} shows that our model fits the $v_c(R)$ data. 
Given that $v_c(R)$ alone can hardly constrain $q$, 
our result is a good demonstration of the usefulness of the HVS data.


\vspace*{0.2 cm}
\scriptsize{
{\it Acknowledgment}: 
KH thanks Dr. de Salas for his help on the $v_c(R)$ data. 
MV and KH are supported by NASA-ATP award NNX15AK79G.
This research was started at the KITP workshop 
`Dynamical Models for Stars and Gas in Galaxies in the \Gaia\ Era' 
held at the Kavli Institute for Theoretical Physics. 
This work has made use of data from the European Space Agency (ESA) mission
{\it Gaia}, processed by the {\it Gaia}
Data Processing and Analysis Consortium (DPAC). Funding for the DPAC
has been provided by national institutions, in particular the institutions
participating in the {\it Gaia} Multilateral Agreement. 
}

\vspace*{-0.5 cm}

\normalsize{ }

\end{document}